# Inverse analysis for the identification of temporal and spatial characteristics of a short super-Gaussian laser pulse interacting with a solid plate

**Karol Pietrak[1]\*, Piotr Łapka[1], Małgorzata Kujawińska[2]**

[1]Institute of Heat Engineering, Warsaw University of Technology,
21/25 Nowowiejska St., 00-665 Warsaw, Poland

\*corresponding author, e-mail: karol.pietrak@itc.pw.edu.pl

[2]Institute of Micromechanics and Photonics, Warsaw University of Technology,
8 Sw. A. Boboli St., 02-525 Warsaw, Poland

**Abstract**

In this paper, the results of numerical experiments verifying a novel setup for laser beam profiling are presented. The experimental setup is based on infrared thermography and includes laser beam illuminating a thin metal plate. The method allows to determine four parameters of the short high-power laser pulse, namely the Super-Gaussian profile coefficient, laser power, pulse start time and duration. The unknown parameters are retrieved based on temporal and spatial temperature distributions at the rear side of the illuminated plate. The applied inverse method is based on Levenberg-Marquardt technique and is implemented in the GNU Octave environment. Solutions of the forward problem are obtained numerically, with the aid of three-dimensional transient heat transfer model implemented in the commercial software ANSYS Fluent. The paper presents the results of the sensitivity analysis as well as calibration and verification of the developed inverse algorithm through application of numerically-generated simulated (artificial) experimental data instead of the physical one. Strengths and weaknesses of the applied approach are widely discussed.

# 1. Introduction

Laser beams of high-energy are encountered in material processing [1] and characterization [2], electro-optical systems [3] and weapon technology [4] among other engineering applications. Many aspects of such laser beam interactions with matter are well-described in the work of von Allmen [5]. The presented study focuses on the identification of transient and spatial characteristics of a high-power super-Gaussian laser pulse interacting with a solid specimen.

In the case of a high-power beam formed in an optical system, the component elements of this system undergo heating during laser operation and their optical surfaces may deform changing the beam profile from the desired Gaussian one to super-Gaussian or even more complex form. It means that it is required to check the profile of a working laser beam. In the industry today, typical laser beam profilers include scanning aperture profilers using slits, knife-edges, or pinholes that utilize single large area detectors, or camera-based profilers using CCD's or photodiode arrays. The high sensitivity of camera profilers require the laser light to be reduced in sensitivity by many orders of magnitude using beam sampling or optical attenuation [6]. Recently a new profiling technique that uses Rayleigh scatter from the beam overcomes the power obstacle and allows measurement and monitoring of the beam caustic and determination of $M^2$ parameters of laser beams with power from 1kW-100kW [7]. In many applications where the Gaussian profile is desired, $M^2$ describes the relative characteristics of the beam and is determined by making multiple measurements of the beam width. However, this instrumentation is very expensive and requires great care in its usage during significant possible changes of a beam parameters. Recently the new approach to this problem was proposed by Kujawińska et al. [8] which is very simple and therefore may be easily applied in the industrial or field conditions with a relatively low cost compared to the other methods. This method assumes that the characteristics of the laser pulse may be found based on temperature distributions recorded with high-speed infrared camera at the rear surface of the heated aluminum plate. The rear surface was selected for collecting the data in order to mitigate the risk of damaging the camera sensors by high-power laser beam which might be reflected from the front surface of the sample. Additional refining of estimated parameter values based on maps of displacements acquired with the aid of Digital Image Correlation (DIC) method [9] is also planned. The DIC method allows to track sample deformation resulting from thermal stresses induced by significant sample heating by the laser pulse [8]. Nevertheless, the current paper is focused only on the details of the thermal part of the introduced problem. It means that the measurement of displacements is not included in the considered inverse method at this stage and only temperature measurements are taken into account in the estimation of laser pulse parameters. The solution of the extended problem, including the finite element displacement model and its incorporation into the inverse estimation procedure will be discussed in a separate article.

Identifications of the unknown surface heat flux based on measured temperatures were discussed before by various authors. General solution approaches for inverse heat transfer problems (IHTPs) employing analytical and numerical heat transfer models were given by Ozisik and Orlande [10] and Beck et al. [11]. Among other notable publications is the one by Huang and Wang [12], who demonstrated the calculation of an unknown boundary heat flux in three-dimensional (3-D) transient inverse heat conduction problem by applying the conjugate gradient method (CGM) and the general purpose commercial software CFX4.2. They assumed infrared temperature detection on one side of a planar body and transient heat flux boundary condition of unknown functional form on the other side, with all surfaces remaining insulated, except for the heated one. In the next work Yang et al. [13]

considered simultaneous estimation of the laser heat flux and melted depth during laser processing in one-dimensional (1-D) geometry. In the research published by Zhou et al. [14] 3-D body subjected to the Gaussian laser beam was considered, with the sinusoidal moderation of the beam intensity. Beam parameters were retrieved based on temperature and heat flux measurements on the back surface with the aid of the CGM. A non-iterative and non-sequential numerical method for solving one- and multi-dimensional transient IHTPs, based on control volume approach, was proposed by Taler and Zima [15]. The method was successfully applied to reconstruct surface heat fluxes in 1-D and 2-D slabs based on temperature measurements. Unfortunately, the proposed approach requires a custom in-house computational model for the solution of thermal problem, which development may be quite cumbersome and time-consuming, especially for heat transfer in the body with complex geometry and a higher number of spatial dimensions. Taler and Taler [16] discussed in detail the intricacies of heat flux and heat transfer coefficient measurement based on temperatures, including design of sensors, their arrangement in various real-life cases, mathematical description of resulting inverse problems and techniques for their solution. In the next work [17] they considered measurements of constant and time-varying heat fluxes or heat transfer coefficients on the surface of a semi-infinite body (1D problems) based on surface temperature measurements. The analysis of uncertainty of the obtained results was performed using the variance propagation rule developed by Gauss. In succession Cebula and Taler [18] developed a space-marching method for determination of transient heat flux distribution on the solid surface based on temperature measurements at selected points located inside the solid. The method was designed for measurements involving surfaces which are inaccessible from the outside, for example the surface of a control rod, and subjected to fast-varying heat fluxes. Similar study by Cebula et al. [19] concerned the measurement of both heat flux and temperature on a cylindrical surface based on the finite element-finite volume method (FEM-FVM) and using thermocouples placed inside the cylinder. Special emphasis was put on the robustness of the method to measurement errors, for example these caused by mispositioned temperature sensors. It is also worthwhile to consider Taler's paper [20] in which a method for determining space-variable heat transfer coefficient using the Levenberg-Marquardt (LM) approach and singular value decomposition (SVD) was presented.

The approach undertaken in the present research is different than the methods discussed above. Firstly, it is aimed at utilization of ready-to-use, well-developed and tested numerical suite for the solution of a direct heat and fluid flow problems (ANSYS Fluent) with its advance functionalities (User Defined Functions, UDF), while many of the published approaches to IHTPs involve analytical models or custom in-house numerical codes. Therefore, presented method is better suited for applied thermal engineering, where bodies of complex geometry are usually encountered. Such geometries can be easily imported to the ANSYS Workbench environment from specialized CAD/CAE programs and then utilized in the numerical simulations. Moreover, application of a solver with multi-physics capabilities allows to easily increase the number of phenomena incorporated in the numerical modeling, which is another advantage of the proposed approach. Secondly, the approach proposed here employs powerful mathematics-oriented free software (GNU Octave) with built-in plotting and visualization tools to develop the proposed overall inverse algorithm. The GNU Octave allows for external solution of the forward problem (ANSYS Fluent), modification of input parameters to the thermal model (modification of input files to the ANSYS Fluent which are read by UDF macros), utilization of the output parameters from the thermal model (reading of output files from ANSYS Fluent which are written by UDF macros), development of the inverse problem solution method by applying build-in libraries, iterative execution of the overall inverse algorithm and analysis of the obtained results. It should be noted that similar approach to the IHTP was demonstrated earlier by Styczniewicz et al. [21, 22] who determined the out-of-plane thermal diffusivity of a thin graphite layer deposited onto a substrate of known thermal properties by means of the flash technique [14, 15]. In their case, the forward problem was solved using the commercial multi-physics software COMSOL and the LM algorithm, applied to find unknown parameters, was implemented in the MATLAB environment. Apart from the out-of-plane thermal diffusivity, the problem involved identification of two other parameters, i.e., surface heat flux and heat transfer coefficient. In the case presented in this paper, the functional form of the recreated boundary heat flux is assumed to be known. Due to that, the function-estimation problem is substituted with a parameter-estimation one, with four unknown parameters, i.e., laser power, dimensionless spatial profile coefficient, start time of the pulse and end time of the pulse. Limited number of unknown parameters allowed for graphic representation of the sensitivity analysis. The usability of the applied method was verified applying numerically-generated simulated (artificial) experimental data instead of data from actual physical experiments.

## 2. Considered inverse problem

The undertaken problem belongs to the class of the IHTPs. In general, they can be viewed as optimization problems in which the sum $\mathbf{S}$ of squared residuals $\mathbf{r}$ is minimized [10]. The sum can be written in matrix notation as:

$$\mathbf{S} = \mathbf{r}^T \mathbf{r} \tag{1}$$

where the residual $\mathbf{r}$ is simply the difference between temperatures predicted by the model of the considered problem $\mathbf{T^m}$ and those obtained experimentally $\mathbf{T^e}$:

$$\mathbf{r} = \mathbf{T^m}(\mathbf{q}) - \mathbf{T^e} \tag{2}$$

Here, the objective function $\mathbf{S}$ depends on some unknown parameters $\mathbf{q}$, as the modeled temperatures $\mathbf{T^m}$ depend on them. The goal of the optimization is to find the set of parameters $\mathbf{q}$ that minimizes $\mathbf{S(q)}$.

Beck and Woodbury [23] gave a general overview of the IHTPs and pointed out specific difficulties characteristic for this class of engineering tasks. Inherent in these problems are following unfavorable properties:

a) possibility of solution non-uniqueness (different vectors of input parameters **q** may result in the same vector of measured quantities – here the same temperatures),

b) ill-conditioning (measured quantities depend weakly on the input parameters, which makes the inversion of the problem difficult – here measured temporal and spatial variations of temperature may weakly depend on laser beam parameters),

c) amplification of measurement and numerical errors.

The inconvenience of type (a) is alleviated by choosing a right initial guess of the unknown parameters for the iterative inverse algorithm to start the search with. As there can be a few local minima, the initial guess that is sufficiently close to the minimum considered in the given problem should be supplied. Type (b) and (c) difficulties are resolved using regularization techniques [23]. In case of the presented method, the LM procedure (extensively characterized in [10]) was applied, in which regularization is obtained by introducing an adaptive damping factor to the gradient-based iterative search.

The important part of the inverse problem solution concerns on the analysis of sensitivity. The measurement technique of the IR thermography results in complete surface fields of temperatures $T(x,y,t)$. The interesting question here is following: in which locations on the sample and at which time moments, the measured quantities are most sensitive to the changes of unknown parameters. The answer can be given if an appropriately detailed model of the problem is available, which may be tested by perturbing relevant parameters and examining resulting variations in measured fields. The sensitivity analysis should be done prior to the physical experiments as it shows when and where the measurements should be taken to minimize the strength of obstacle of type (b). Such analysis is carried out in this paper.

## 3. Problem statement

**3.1 Direct problem**

The problem involved solid plate made of aluminum alloy AW2017A T4 as presented in Fig. 1. The initial studies revealed that thinner specimens were preferred than thicker ones due to registration of the temperature on the opposite surface (rear) to the irradiated (front) one. Therefore, the aluminum sample was made thinner in the central part by circular milling. Parallel planed experimental measurements [8] assume non-destructive character of the tests with expected temperature levels in the specimen below the melting temperature for aluminum ($T_m$ = 883.0 K). Therefore, the melting process was not accounted for. Thermophysical properties as well as geometry of the specimen are specified in Tab. 1. The material properties were assumed to be isotropic. Moreover, the thermal conductivity and specific heat were temperature dependent [8] – see Tab. 1. Taking into account assumptions listed above, the energy equation in the sample took the following form:

$$\rho c_p(T)\frac{\partial T}{\partial t} = \frac{\partial}{\partial x_i}\left[\lambda(T)\frac{\partial T}{\partial x_i}\right] \qquad (3)$$

where the subscript $i = 1, 2$ and $3$ denotes $x$-, $y$- and $z$-axis of the Cartesian coordinate system, the symbol $c_p$ is the specific heat, $T$ – temperature, $t$ – time, $\lambda$ – thermal conductivity and $\rho$ – density.

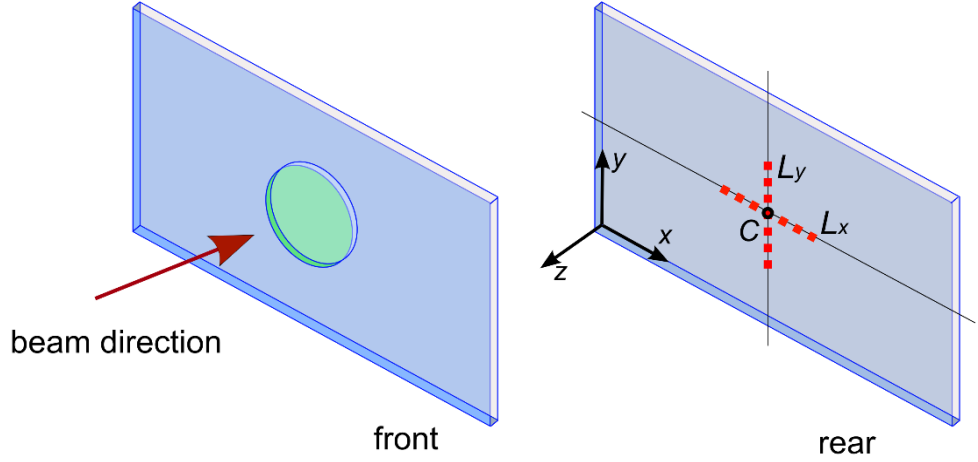

Fig. 1. The geometry of analyzed sample and locations of temperature measurements. Temperature history data from sections $L_x$ and $L_y$ were used to recreate the parameters of super-Gaussian laser pulse. The temporal data from point $C$ denoting the intersection of $L_x$ and $L_y$ was also used in the inverse analysis.

Tab. 1. Geometrical and thermophysical properties of the considered sample

| **Dimensions (width×length×thickness)** | $a\times b\times h$ [mm] | 80.0×50.0×1.0 | | | | | | |
|---|---|---|---|---|---|---|---|---|
| **Milling** | $R_m\times\delta$ [mm] | 2.0×0.5 | | | | | | |
| **Density** | $\rho$ [kgm$^{-3}$] | 2700.0 | | | | | | |
| | [°C] | 30 | 50 | 100 | 150 | 200 | 250 | 300 |
| **Thermal conductivity** | $\lambda$ [Wm$^{-1}$K$^{-1}$] | 122.6 | 125.0 | 133.7 | 151.9 | 154.9 | 130.0 | 159.6 |
| **Specific heat** | $c_p$ [kJkg$^{-1}$K$^{-1}$] | 0.902 | 0.911 | 0.943 | 1.046 | 1.039 | 0.832 | 0.995 |

The initial temperature field in the sample was assumed uniform and equal to the surroundings temperature, i.e., $T_0 = T_\infty = 300$ K. The boundary condition for the heated wall of the sample (front wall) was following:

$$-\lambda(T)\frac{\partial T}{\partial n}\bigg|_w = q_{laser} + h(T_\infty - T_w) + (1-r)\sigma(T_\infty^4 - T_w^4) \qquad (4)$$

where the subscript $w$ denotes the sample wall, $\infty$ – surroundings, the symbol $h$ is the heat transfer coefficient (assumed value was $h = 5$ Wm$^{-2}$K, which is the typical for natural convection in the case of horizontally oriented plates [24]), $n$ – normal direction to the wall, $r$ – surface reflectivity (assumed value was $r = 0.7$) and $\sigma$ – Stefan-Boltzmann constant ($\sigma = 5.67\cdot 10^{-8}$ Wm$^{-2}$K$^{-4}$). The following equation described the heat flux related to the laser pulse which incident perpendicularly to the heated surface [25]:

$$q_{laser} = \begin{cases} (1-r)Q \dfrac{p 2^{\frac{2}{p}}}{2\pi R_0^2 \Gamma\left(\frac{2}{p}\right)} \exp\left[-2\left(\dfrac{R}{R_0}\right)^p\right] & \text{for } t_{start} \leq t \leq t_{end} \\ 0.0 & \text{for } 0 \leq t < t_{start} \text{ and } t > t_{end} \end{cases} \quad (5)$$

where the symbol $p$ is dimensionless shape coefficient of the super-Gauss function, $Q$ – laser power, $R$ – radial distance from the beam center to given location, $R_0$ – the length over which the Gaussian profile (for $p = 2$) decreases to $e^{-2}$ of its axial value (assumed value was $R_0 = 0.005$ m), $t_{start}$ – time of the start of the laser pulse, $t_{end}$ – time of the end of the laser pulse and $\Gamma$ – Euler gamma function. The temporal profile of the excitation is assumed rectangular, i.e., the effects of power increase and decrease are neglected. Such simplification is justifiable considering the discussion presented by McMasters and Dinwiddie [26] and the fact that the duration of the utilized laser pulses is relatively long (of the order of milliseconds).

At rear wall the radiative and convective heat loss were accounted for, therefore the boundary condition was in the following form:

$$-\lambda(T)\dfrac{\partial T}{\partial n}\bigg|_w = h(T_\infty - T_w) + (1-r)\sigma(T_\infty^4 - T_w^4) \quad (6)$$

while lateral walls were treated as adiabatic, hence:

$$-\lambda(T)\dfrac{\partial T}{\partial n}\bigg|_w = 0 \quad (7)$$

The solution of the forward problem was obtained numerically via the finite volume method. The solver of choice was the ANSYS Fluent 17.2. The UDFs were applied to model temporary and spatially variable surface heat flux resulting from the short high-power laser pulse interaction with the solid body. The geometry of the sample was prepared in the ANSYS DesignModeler 17.2, while the grid was generated in the ANSYS Meshing 17.2 using sweep method and consisted of almost 198 thousand of hexahedron and wedge elements. The spatial and temporal discretization validity was checked by conducting a series of model evaluations. The selected temporal and spatial discretization parameters ensured solutions which were time stable and independent on the time step and grid element sizes. The mesh quality parameters were as follows: skewness below 0.7 and orthogonal quality above 0.3.

### 3.2 Inverse problem

In the heat transfer problem formulated above, 4 unknown parameters were assumed, that is:
1) laser power $Q$,
2) dimensionless shape coefficient $p$,
3) start time of the laser pulse $t_{start}$,

4) end time of the laser pulse $t_{end}$.

Time and spatial variations of temperatures at the rear surface of the sample (oppose to the heated one) were assumed available (e.g., from simulated numerical or physical experiments). The locations of measurement points were limited to two sections denoted by $L_x$ and $L_y$ and placed at symmetry planes of the specimen, directly vis-à-vis the illuminated area at front surface, as shown in Fig. 1. Such locations were selected based on preliminary numerical analyses of temperature and displacement fields presented in [8]. They indicated that temperature distributions are axisymmetric in relation to $z$-axis, while displacements have two planes of symmetry, i.e., $x$ and $y$.

## 4. Solution method

**4.1 Iterative procedure**

The inverse problem stated above was solved using the LM algorithm, as described by Ozisik and Orlande [10]. The general schematic of the solution methodology is presented in Fig. 2. The developed procedure was implemented in the GNU Octave environment. In the method proposed in [10], the unknown parameters were found iteratively, starting from the supplied initial values. The values of unknown parameters for iteration $j+1$ were calculated based on data from the previous iteration $j$ in accordance with the following equation:

$$\mathbf{q}^{j+1} = \mathbf{q}^j + \left[\mathbf{J}^T(\mathbf{q}^j) \cdot \mathbf{J}(\mathbf{q}^j) + \mu^j \mathbf{\Omega}\right]^{-1} \cdot \mathbf{J}^T(\mathbf{q}^j)\left[\mathbf{T}^e - \mathbf{T}^m(\mathbf{q}^j)\right] \tag{8}$$

where the symbol $\mathbf{J}(\mathbf{q}^j)$ is the sensitivity (Jacobian) matrix, $\mu^j$ – adaptive regularization coefficient and $\mathbf{\Omega}$ – diagonal amplification matrix. All these quantities are explained below.

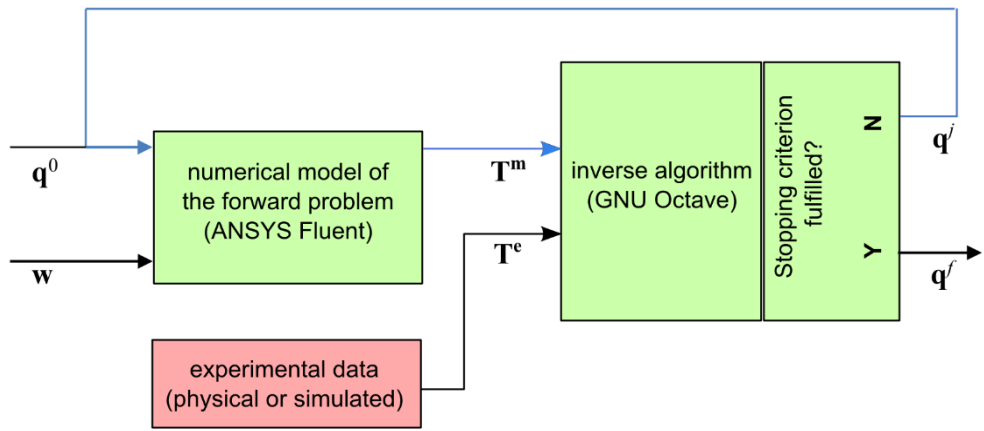

Fig. 2. Schematic of the solution methodology, where the symbol $\mathbf{q}^0$ is the initial guess of the unknown system parameters, $\mathbf{w}$ – known system parameters, $\mathbf{T}^e$ – experimental values of temperature, $\mathbf{T}^m$ – modeled temperature values, $\mathbf{q}^j$ – unknown parameters estimated in $j$-th iteration, $\mathbf{q}^f$ – final values of estimated parameters.

The elements of the Jacobian matrix $J_{ik}$ (where the subscript $i$ is the index of the element of temperature vectors $\mathbf{T^e}$ and $\mathbf{T^m}$ and $k \in 1 \div 4$ – index of the unknown parameter) were approximated by the finite differences, according to the following formula:

$$J_{ik} = \frac{\partial T_i^m}{\partial q_k} \cong \frac{T_i^m(q_k + \Delta q_k) - T_i^m(q_k)}{\Delta q_k} \quad (9)$$

In eq. (9) the symbol $\Delta q_k$ denotes the perturbation of $k$-th unknown parameter. In practice, single computation of the sensitivity matrix required simulating the numerical model 5 times – one time with the current values of unknown parameters (to obtain unperturbed temperatures $T_i^m(q_k)$), and 4 times with perturbed value of each unknown parameter separately (to obtain perturbed temperatures $T_i^m(q_k + \Delta q_k)$). The disturbances $\Delta q_k$ were chosen as 1/10 of the initial guess vector $\mathbf{q}^0$ defined in the following way:

$$\mathbf{q}^0 = [q_1^0, q_2^0, q_2^0, q_4^0]^T \quad (10)$$

where symbols $q_1^0$, $q_2^0$, $q_3^0$ and $q_4^0$ are the laser power, laser pulse dimensionless shape coefficient, start time of the laser pulse and end time of the laser pulse, respectively.

The matrix $\mathbf{\Omega}$ was assumed after Marquardt [27] as:

$$\mathbf{\Omega} = \mathbf{J}^T(\mathbf{q}^j) \cdot \mathbf{J}(\mathbf{q}^j) \quad (11)$$

whereas the starting value of the scalar $\mu$ was found by experimentation and was set to $\mu^0 = 0.5$. This scalar was modified in each iteration of the developed LM-based inverse procedure. On one hand, too small value of $\mu$ at the beginning of the iterative search may result in large oscillations in computed values of the unknown parameters – the procedure is then unstable. On the other hand $\mu$ may be decreased when the search arrives near the minimum of the objective function (see eq. (1)) to allow for fine adjustments of the values of the unknown parameters. The developed procedure decreased $\mu$ by the factor of 0.3 if the sum $\mathbf{S}^{j+1}$ for the current iteration was smaller or equal to the sum $\mathbf{S}^j$ for the previous iteration. Otherwise $\mu$ was increased by the factor of $0.3^{-1}$.

**4.2 Stopping criteria**

There are three commonly-used criteria that allow to accept parameters, calculated in $j$-th iteration of the LM method as the final solution of the posed inverse problem. In the mathematical notation they may be written as follows:

$$\mathbf{S}(\mathbf{q}^{j+1}) < \varepsilon_1 \quad (12)$$

$$\|\mathbf{J}^T(\mathbf{q}^j)[\mathbf{T^e} - \mathbf{T^m}(\mathbf{q}^j)]\| < \varepsilon_2 \quad (13)$$

$$\|\mathbf{q}^{j+1} - \mathbf{q}^j\| < \varepsilon_3 \quad (14)$$

where the constants $\varepsilon_1$, $\varepsilon_2$ and $\varepsilon_3$ are experimentally-determined scalars and the symbol $\|.\|$ denotes Euclidean norm of a vector. The first criterion given by eq. (12) uses the fact that the least squares norm is minimal at optimal solution. The second criterion described by eq. (13) is based on the observation that the gradient vanishes at the global minimum (this is also true for local minima and saddle points of the objective function, yet the LM-based inverse methods are unlikely to converge to such points). The last criterion, i.e., inequality given by eq. (14) expresses the fact, that near the global minimum, the parameters calculated in the consecutive iterations are very similar. In this work selection of values of the constants $\varepsilon_1$, $\varepsilon_2$ and $\varepsilon_3$ was done after several test runs of the developed inverse procedure. They were set to $\varepsilon_1 = 35$, $\varepsilon_2 = 5$ and $\varepsilon_3 = 0.045$ (determined for power in kW, temporal parameters in ms and least squares norm **S** defined in section 5.2).

# 5. Results and discussion

**5.1 Sensitivity analysis**

The experimental measurements of temporal and spatial variation of the temperature on the rear side of the sample are planned to be carried out applying high-speed infrared camera, e.g., FLIR SC7500 IR. The frame rate of this camera is 1250 Hz at 160×128 pixels, which corresponds to a time step size of $\Delta t = 0.0008$s. The area of detection may range over approximately 20×20 pixels. In the simulated numerical experiments performed in this paper the frames collected at all time instants were assumed available, but for the sake of simplicity it was decided to limit the temperature data used by the inverse procedure to three sets of data, i.e., $\mathbf{T_t}$, $\mathbf{T_x}$ and $\mathbf{T_y}$. For an accurate retrieval of temporal parameters ($t_{start}$ and $t_{end}$) temperature variation at point $C$ placed at the intersection of $L_x$ and $L_y$ (see Fig. 1.) was used. This temporal distribution of data was denoted as $\mathbf{T_t}$. Identification of the $p$-parameter, which describes the flatness of the super-Gaussian profile, required the spatial temperature profiles along sections $L_x$ and $L_y$ to be included in the measurement vector $\mathbf{T^e}$. These spatial distributions of data were denoted as $\mathbf{T_x}$ and $\mathbf{T_y}$. In general, these profiles may be taken from one or more time instants (frames), preferably near the moment when the temperature increase on the detector side of the sample approaches maximum. For the sake of clarity of the sensitivity analysis presented in this paper, the spatial temperature profiles from a single moment in time ($t = t_{start}+2$ ms) were considered. The measurement vector for sensitivity analysis was defined as a superposition of three aforementioned vectors, which may be written as follows:

$$\mathbf{T^e} = \left[\mathbf{T_x^e}, \mathbf{T_y^e}, \mathbf{T_t^e}\right]^T \quad (15)$$

The vector of modeled temperatures $\mathbf{T^m}$ was constructed to correspond with the experimental vector shown above and was following:

$$\mathbf{T^m} = \left[\mathbf{T_x^m}, \mathbf{T_y^m}, \mathbf{T_t^m}\right]^T \quad (16)$$

The magnitudes of coefficients $J_{ik}$ of the Jacobian matrix defined by eq. (9) indicate the sensitivity of the elements of temperature vector $\mathbf{T^m}$ to changes of unknown parameters 1-4 in the vector $\mathbf{q}$. On sensitivity diagrams (Fig. 3) one can see the values of modeled temperatures (components of vector $\mathbf{T^m}$) as functions of space and time for the following initial guess: $q_1^0 = 5$ kW, $q_2^0 = 5$, $q_3^0 = 1$ ms and $q_4^0 = 3$ ms. The values of numerically-calculated sensitivity coefficients $J_{ik}$ corresponding to the aforementioned initial guess vector are also shown in Fig. 3, directly below the temperature vector components.

Most important conclusion from the analysis of sensitivities was that temperature acquisition should last long enough to include the period in which the rear surface of the sample begins to cool down in the surrounding. The sensitivity of the system to changes of parameter no. 4 in vector **q** ($t_{end}$) was nonzero only after the end of thermal excitation (laser pulse), and its maximal value was not reached immediately but with some delay.

**5.2 Results of numerical simulations**

The conclusions from the sensitivity analysis helped in designing simulated experiments in which data were not taken from physical experiments, but generated with the use of the numerical model. In such case the exact values of parameters $\mathbf{q^e}$ were known, and the performance of the iterative procedure might be easily assessed. As real thermograms are characterized by the presence of noise, uniformly-distributed pseudorandom noise of peak amplitudes ±1 K, ±2 K and ±3 K and with zero expected value was added to the simulated experimental temperatures $\mathbf{T^e}$. Such treatment allowed to verify the robustness of proposed method to measurement errors.

It was decided that the inverse procedure should use spatial temperature distribution data ($\mathbf{T_x}$ and $\mathbf{T_y}$) from three time instants (frames) instead of from one as was done in the sensitivity analysis. First of the utilized frames was collected just before the laser pulse ended, while the two other frames were collected with a delay of 1.6 and 3.2 ms in relation to the first frame, respectively (see Fig. 4 (c)). The temporal variation of temperature $\mathbf{T_t}$ in point *C* (Fig. 1) was also used. The data from these three frames and temperature history in point *C* were included in the minimized least squares norm. The information about temporal variation of temperature at point *C* ($\mathbf{T_t}$) was repeated twice in the vectors $\mathbf{T^e}$ and $\mathbf{T^m}$. Such treatment improved convergence of temporal and power parameters.

The results of numerical experiments conducted with these modifications were grouped in Tab. 2 and Tab. 3. Tab. 2 shows individual results of benchmark tests for noisy data of peak noise amplitude ±1 K, but simulations using identical settings were also carried out for data with greater noise amplitudes and their cumulative results are shown in Tab. 3, where the mean absolute percentage error (MAPE) is calculated for each considered parameter separately as well as cumulatively for all of them using the following formula:

$$\text{MAPE} = \frac{100}{N} \sum_{i=1}^{N} \frac{|\Delta q_i|}{q_i^e} \% \qquad (17)$$

where *N* is the number of estimations, $q^e$ – the accurate value of the unknown parameter, $\Delta q$ – the difference between the estimated and the accurate value and |.| denotes the absolute value.

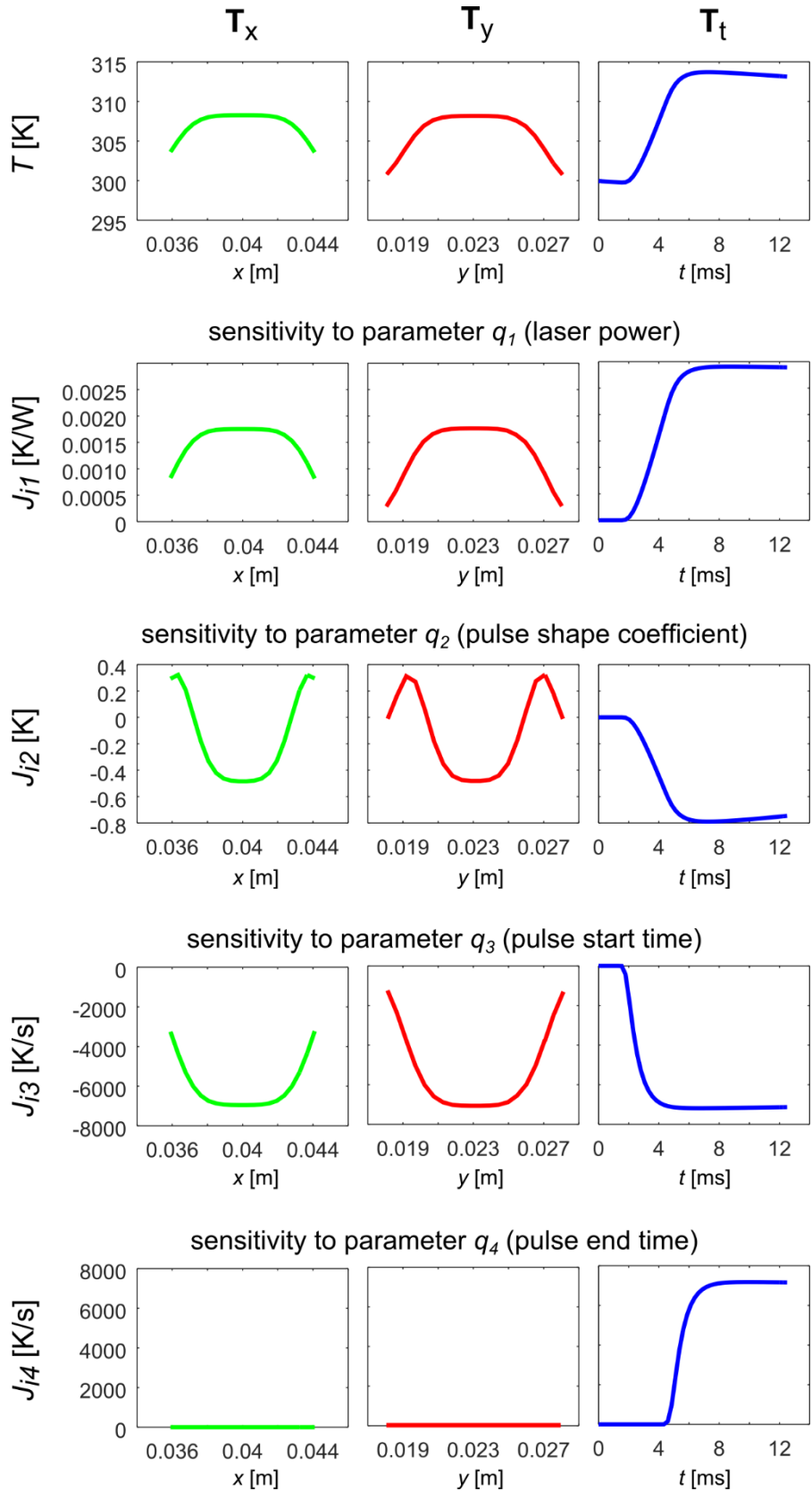

Fig. 3. Values of modeled temperatures $\mathbf{T^m}$ for parameter vector $\mathbf{q}^0 = [5\ (\text{kW}), 5,\ 1\ (\text{ms}), 3\ (\text{ms})]^T$ and their sensitivity to perturbations of unknown parameters 1-4 in the vector $\mathbf{q}$, where: $q_1$ corresponded to laser power $Q$, $q_2$ – dimensionless spatial laser pulse shape coefficient $p$, $q_3$ – time of the laser pulse start $t_{start}$ and $q_4$ – end time of the laser pulse $t_{end}$. The diagrams show values of coefficients in each column of sensitivity matrix, respectively.

Tab. 2. The results of numerical experiments for retrieval of parameters based on noisy data (peak amplitude of noise ±1 K). The symbols $q^e$ and $q^m$ denote desired and obtained values of unknown parameters, respectively. Absolute value is denoted by the symbol |.|.

| Exp. no. | quantity | $Q$ [kW] | $p$ [-] | $t_{start}$ [ms] | $t_{end}$ [ms] |
|---|---|---|---|---|---|
| 1 | $q_i^e$ | **10** | **2** | **2** | **4** |
| 1 | $q_i^m$ | **9.906** | **2.038** | **1.994** | **4.032** |
| 1 | $|\Delta q_i|$ | 0.094 | 0.038 | 0.006 | 0.032 |
| 1 | $|\Delta q_i/q_i^e|$ | 0.94% | 1.91% | 0.30% | 0.79% |
| 2 | $q_i^e$ | **10** | **2** | **3** | **5** |
| 2 | $q_i^m$ | **9.521** | **2.039** | **2.928** | **5.069** |
| 2 | $|\Delta q_i|$ | 0.479 | 0.039 | 0.072 | 0.069 |
| 2 | $|\Delta q_i/q_i^e|$ | 4.79% | 1.95% | 2.42% | 1.37% |
| 3 | $q_i^e$ | **10** | **12** | **2** | **4** |
| 3 | $q_i^m$ | **10.112** | **11.917** | **2.000** | **4.018** |
| 3 | $|\Delta q_i|$ | 0.112 | 0.083 | 0.000 | 0.018 |
| 3 | $|\Delta q_i/q_i^e|$ | 1.12% | 0.70% | 0.00% | 0.44% |
| 4 | $q_i^e$ | **10** | **12** | **3** | **5** |
| 4 | $q_i^m$ | **9.655** | **10.988** | **2.946** | **5.050** |
| 4 | $|\Delta q_i|$ | 0.345 | 1.012 | 0.054 | 0.050 |
| 4 | $|\Delta q_i/q_i^e|$ | 3.45% | 8.43% | 1.80% | 1.00% |
| 5 | $q_i^e$ | **20** | **2** | **2** | **4** |
| 5 | $q_i^m$ | **19.976** | **2.003** | **1.997** | **4.038** |
| 5 | $|\Delta q_i|$ | 0.024 | 0.003 | 0.003 | 0.038 |
| 5 | $|\Delta q_i/q_i^e|$ | 0.12% | 0.17% | 0.14% | 0.94% |
| 6 | $q_i^e$ | **20** | **2** | **3** | **5** |
| 6 | $q_i^m$ | **19.571** | **2.003** | **2.956** | **5.069** |
| 6 | $|\Delta q_i|$ | 0.429 | 0.003 | 0.044 | 0.069 |
| 6 | $|\Delta q_i/q_i^e|$ | 2.15% | 0.16% | 1.47% | 1.38% |
| 7 | $q_i^e$ | **20** | **12** | **2** | **4** |
| 7 | $q_i^m$ | 19.479 | 12.181 | 1.927 | 4.081 |
| 7 | $|\Delta q_i|$ | 0.521 | 0.181 | 0.073 | 0.081 |
| 7 | $|\Delta q_i/q_i^e|$ | 2.61% | 1.51% | 3.66% | 2.01% |
| 8 | $q_i^e$ | **20** | **12** | **3** | **5** |
| 8 | $q_i^m$ | **19.928** | **11.851** | **3.002** | **4.983** |
| 8 | $|\Delta q_i|$ | 0.071 | 0.149 | 0.003 | 0.017 |
| 8 | $|\Delta q_i/q_i^e|$ | 0.36% | 1.24% | 0.09% | 0.34% |

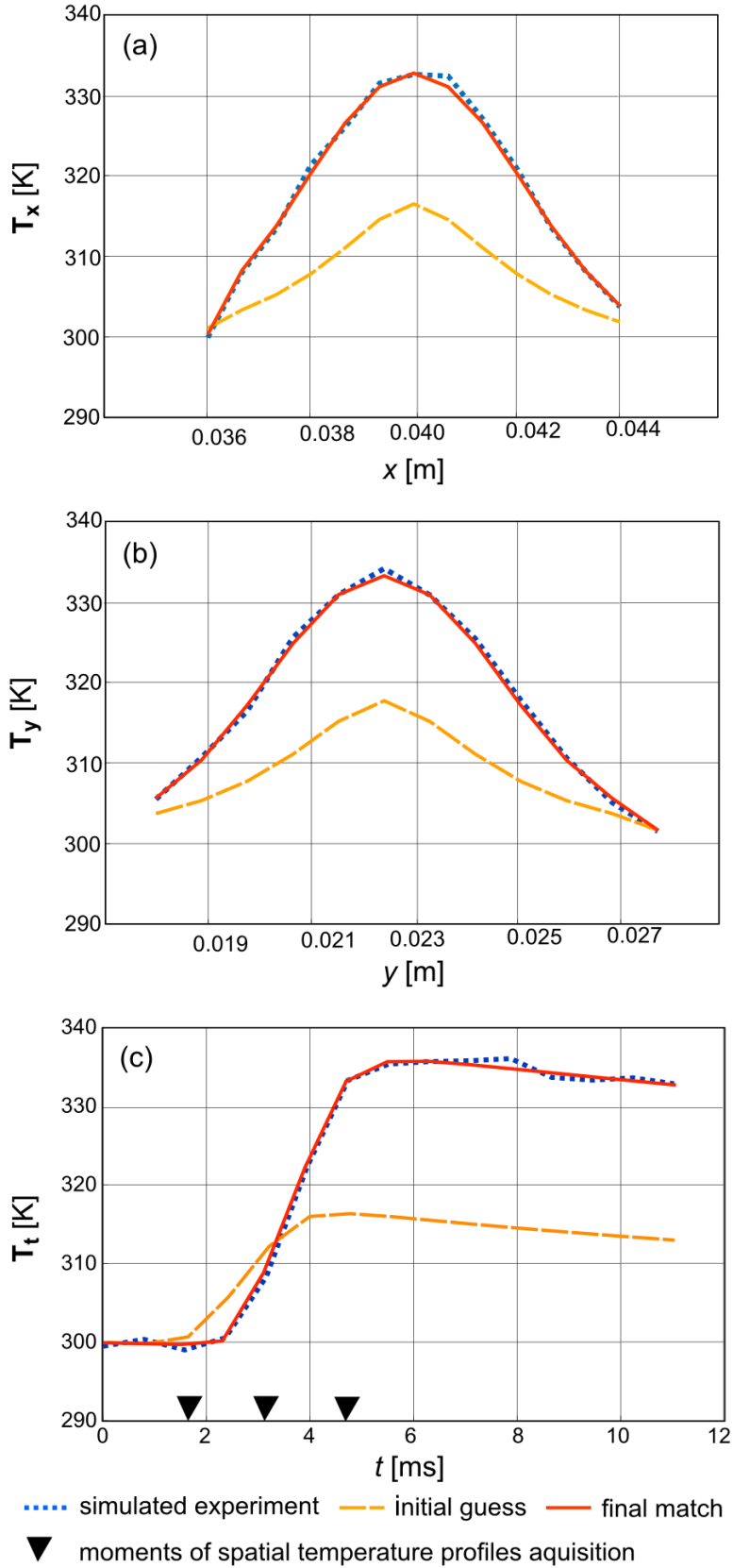

Fig. 4. An example of fitted values of simulated experimental temperatures $T^e$ (dotted lines) and modeled temperatures $T^m$ (solid lines) for experiment no. 1 from Tab. 2: a) and b) spatial distributions of temperature along $L_x$ and $L_y$, respectively for last of the utilized time instants and c) temporal temperature variation in point $C$. Temperatures generated as initial guess are also shown (dashed lines). Please note that the inverse algorithm used $T_x$ and $T_y$ data from three time instants marked by black triangles in diagram (c).

Tab. 3. Mean absolute percentage errors (MAPEs) obtained in numerical retrieval experiments for all considered unknown parameters for different peak amplitudes of applied pseudorandom noise

| Noise peak amplitude | MAPE for all parameters | MAPE($Q$) | MAPE($p$) | MAPE($t_{start}$) | MAPE($t_{end}$) |
|---|---|---|---|---|---|
| ±1 K | 1.55% | 1.94% | 2.01% | 1.23% | 1.04% |
| ±2 K | 3.39% | 5.39% | 3.98% | 2.55% | 1.79% |
| ±3 K | 3.39% | 5.99% | 3.53% | 2.68% | 1.36% |

For each peak noise amplitude, testing with both flat-top ($p = 12$) and round-top ($p = 2$) beams, for two different laser powers ($Q = 10$ and $Q = 20$ kW) and two different temporal characteristics of the pulse ($t_{start} = 2$ ms and $t_{end} = 4$ ms, as well as $t_{start} = 3$ ms and $t_{end} = 5$ ms) was performed. The main goal was that the algorithm should arrive at sufficiently accurate values of unknown parameters (within ±5% from the numerically-generated simulated experimental values) in a reasonable number of iterations (maximal allowed number of iterations was set to 15), starting from the same initial guess in each simulation, i.e., the starting vector was $q_1^0 = 2$ kW, $q_2^0 = 1.2$, $q_3^0 = 1$ ms and $q_4^0 = 3$ ms. All together 24 numerical experiments were carried out – eight at each noise level. In cases of lower noise level (± 1 K), the algorithm stopped due to criterion given by eq. (12) or (14) after approximately 6-7 iterations. For two greater noise levels (± 2 and ±3 K), it stopped due to criterion described by eq. (14) or after reaching the maximal number of iterations. An example of fitted data for experiment no. 1 from Tab. 2 is shown in Fig. 4. In Fig. 4(a) and (b) one can see the spatial distributions of temperature for initial guess, simulated experiments and final fit along $L_x$ and $L_y$ from the last of the utilized time instants while Fig. 4(c) presents temporal temperature variation in point $C$. The spatial distributions of temperature from two earlier time instants are not depicted.

### 5.3 Discussion

The presented procedure proved to be effective in the undertaken task. The MAPEs calculated for all parameters were 1.55 %, 3.39 % and 3.39 % for noise levels ±1, ±2 and ±3 K, respectively. Standard deviations from the target values were computed for each parameter separately. When using the highest noise level, they were: 1.3117 kW, 0.3537, 0.0783 ms and 0.0887 ms for the laser power, pulse shape coefficient, start time and end time, respectively. The difference between estimated and expected values in some cases exceeded 5%, e.g., the error related to recovering the parameter $p$ in the experiment no. 4 in Tab. 2 reached 8.43%. Nevertheless, in most cases, accurate values were followed with good accuracy, and the cumulative mean values of errors were satisfactory.

Main disadvantages of the developed method are iterative character and numerical computation of the sensitivity matrix at each iteration of the solution procedure (the latter is necessary for systems described by nonlinear equations as their Jacobian depends on the unknown parameters, contrary to linear systems where the sensitivity coefficients are constant). The finite difference approximation of the Jacobian required $N+1$ model evaluations, where $N$ was the number of unknowns. That makes computations relatively expensive for models with fine discretization and high number of unknowns. Fortunately, the problem of available computational resources is largely reduced nowadays, and the simulations carried out in this paper were handled by applying parallel processing approach for the solution of the forward problem, which makes computation times acceptable.

From papers [16–19] it can be deducted that most accurate identification of fast-varying heat fluxes on solid surfaces is assured if temperature sensing is carried out directly from the heated surface or as close to it as possible. The main drawback of front-face sensing is the possibility of sensor damage resulting from the reflected laser radiation. This is especially important during dealing with high-power laser pulse. Therefore, it was decided to use rear sensing and to reduce the plate thickness in the active area to prevent temperature damping by the plate.

The case study presented here may be treated as demonstration of a general solution framework suitable for many engineering and scientific problems. In this philosophy, each task is handled by separate application (the inverse analysis – GNU Octave, the forward problem – ANSYS Fluent with UDF macros) which allows to take advantage of the strengths and potentials of both engineering programs. In the problem presented in this paper, the GNU Octave was the master program and the ANSYS Fluent was the slave, invoked with a journal file modified beforehand with the use of an in-house GNU Octave routine. The communication between applications was carried out using file input/output and specially prepared UDF macros.

## 6. Summary

The presented work is summarized in the following points:
1. The inverse problem solution for a method of identification of spatio-temporal characteristics of high-power laser pulse interacting with the aluminum plate was presented. The method takes advantage of a fast infrared camera which collects temperature data from the rear surface of a thin irradiated metal sample. The simplicity of noncontact full-field data capture in combination with the developed numerical method guarantees high experimental applicability in the harsh environmental conditions.
2. The data analysis algorithm, implemented in a multipurpose open-source scientific software GNU Octave, is based on the LM technique and utilizes the well-known ANSYS Fluent solver with UDFs macros for the solution of the forward problem.
3. Sensitivity analysis revealed that sensitivity of the temperature response signal to parameter no. 4 (the end time of the laser pulse) is nonzero only after the signal reaches its maximum and the sample begins to cool down.
4. The accuracy of the method was assessed with the aid of numerically-generated simulated (artificial) experimental data. The method proved to be effective in the undertaken task. The mean absolute percentage errors of estimations as high as 1.55%, 3.39% and 3.39% were obtained for noise levels ±1, ±2 and ±3 K, respectively.

5. Main advantage of the method is the capability to use external solvers to solve the direct problem. Here, application of the GNU Octave scripts and ANSYS Fluent with UDF macros allowed to use advanced CFD solver which allows to simulate many heat transfer phenomena in complex geometries. Presented algorithm is an example of a general framework for inverse problems, applicable to many engineering and scientific problems.

6. The iterative character of the method resulted in the long computation times may be seen as its main weakness.

## 7. Acknowledgements

This work was supported by the National Centre for Research and Development (Poland) under grant no. DOB-16-6/1/PS/2014 and the statutory funds of Faculty of Mechatronics and Faculty of Power and Aeronautical Engineering of Warsaw University of Technology.